\documentclass[
  aps,
  prl,
  twocolumn,
  showpacs,
  superscriptaddress
]{revtex4-2}
\usepackage{graphicx}  
\usepackage{dcolumn}   
\usepackage{bm}        
\usepackage{booktabs}
\usepackage{amsmath,amsfonts,amssymb}
\usepackage{lineno}
\usepackage{physics}
\usepackage{booktabs} 
\usepackage{lipsum}
\usepackage{tcolorbox}
\usepackage{float}
\usepackage{multirow}
\usepackage{xcolor}
\usepackage{comment}
\usepackage{caption}
\usepackage{relsize}
\usepackage[normalem]{ulem}
\usepackage[utf8]{inputenc}
\usepackage{svg}
\usepackage{tikz}
\usetikzlibrary{arrows.meta, positioning}

\usepackage{url}

\newcommand\RNGRate{$21.7\pm0.5$ }

\begin{document}

\title{On-chip semi-device-independent quantum random number generator exploiting contextuality}
\author{Maddalena Genzini}
\affiliation{Department of Electrical and Photonics Engineering, Technical University of Denmark, Ørsteds Pl., Kgs. Lyngby 2800, Denmark}
\author{Caterina Vigliar}
\affiliation{Department of Electrical and  Photonics Engineering, Technical University of Denmark, Ørsteds Pl., Kgs. Lyngby 2800, Denmark}
\author{Mujtaba Zahidy}
\affiliation{Department of Electrical and Photonics Engineering, Technical University of Denmark, Ørsteds Pl., Kgs. Lyngby 2800, Denmark}
\author{Hamid Tebyanian}
\affiliation{School of Physical and Chemical Sciences, Queen Mary University of London, London, United Kingdom}
\author{Andrzej Gajda}
\affiliation{IHP – Leibniz-Institut für innovative Mikroelektronik, 15236 Frankfurt (Oder), Germany}
\author{Klaus Petermann}
\affiliation{Technische Universität Berlin, Institut f{\"u}r Hochfrequenz- und Halbleiter-Systemtechnologien, 10587 Berlin, Germany}
\author{Lars Zimmermann}
\affiliation{IHP – Leibniz-Institut für innovative Mikroelektronik, 15236 Frankfurt (Oder), Germany}
\affiliation{Technische Universität Berlin, FG Silizium-Photonik, 10587 Berlin, Germany}
\author{Davide Bacco}
\affiliation{University of Florence, Department of Physics and Astronomy, 50019 Sesto Fiorentino, Italy}
\affiliation{QTI s.r.l., 50125, Firenze, Italy}
\author{Francesco Da Ros}
\affiliation{Department of  Electrical and Photonics Engineering, Technical University of Denmark, Ørsteds Pl., Kgs. Lyngby 2800, Denmark}

\begin{abstract}
We present a semi-device-independent quantum random number generator (QRNG) based on the violation of a contextuality inequality, implemented by the integration of two silicon photonic chips. Our system combines a heralded single-photon source with a reconfigurable interferometric mesh to implement qutrit state preparation, transformations, and measurements suitable for testing a KCBS contextuality inequality. This architecture enables the generation of random numbers from the intrinsic randomness of single-photon interference in a complex optical network, while simultaneously allowing a quantitative certification of their security without requiring entanglement. We observe a contextuality violation exceeding the classical bound by more than $10\sigma$, unambiguously confirming non-classical behavior. From this violation, we certify a conditional min-entropy per experimental round of $H_{min}=0.077\pm0.002$, derived via a tailored semidefinite-programming-based security analysis. Each measurement outcome therefore contains at least 
$0.077\pm0.002$ bits of extractable genuine randomness, corresponding to an asymptotic generation rate of \RNGRate bits/s. These results establish a viable route towards semi-device-independent quantum random number generators compatible with practical integrated photonic quantum networks.

\end{abstract}

\maketitle
\section{Introduction}

Randomness lies at the center of several important computational applications, such as Monte Carlo simulations~\cite{l1990random,harrison2010introduction}, neural-networks weighting~\cite{makovoz1996random}, and secure communication and data encryption~\cite{gisin2002quantum,l2012random,dengrandom}. With the advance in cryptographic standards and the growing concern in data protection, the demand for true random numbers has become critical. Over the last decade, several schemes have emerged that leverage the non-deterministic nature of quantum mechanics to design quantum random numbers generators (QRNGs)~\cite{jacak2021quantum,vallone2014quantum,aldama2022integrated,nie2015generation,gabriel2010generator,pironio2010random,shalm2021device,liu2018device,PhysRevLett.130.080802,um2013experimental,gehring2021homodyne}.

These schemes can be classified based on the level of trust required for their implementation. Fully trusted ones assume complete trust in the hardware and harness internal quantum processes to generate randomness, such as laser phase noise~\cite{nie2015generation}, vacuum fluctuations~\cite{gabriel2010generator}, or quadrature fluctuations~\cite{gehring2021homodyne}. In practice, the security of such schemes can be jeopardized due to tampering. Device-independent (DI) schemes, on the other hand, assume no trust in the hardware by relying on the violation of Bell’s inequalities~\cite{pironio2010random,shalm2021device,liu2018device}. 
More specifically, the non-local correlations of entangled states can be used to certify the presence of genuine randomness without requiring any assumption about the internal working of the QRNG system. Recent experiments have further demonstrated DI randomness amplification, where imperfect randomness sources are transformed into nearly ideal randomness through Bell-certified quantum correlations~\cite{kulikov2026experimental}. However, their need for high-quality entanglement, efficient photon detection, and space-like separation restricts rates to kb/s per second compared to the range of Gb/s per second of fully trusted approaches~\cite{shalm2021device}. 

Semi-device-independent (semi-DI) schemes have therefore emerged as a practical intermediate regime between these two extremes. By relaxing some assumptions about the devices, while still retaining partial physical constraints such as dimensional bounds or measurement compatibility, semi-DI protocols can certify randomness under weaker trust assumptions than fully trusted QRNGs while remaining experimentally accessible.~\cite{PhysRevLett.130.080802, um2013experimental}.

Contextuality-based approaches use inequalities built upon the Kochen-Specker theorem~\cite{specker1990logik,kochen1990problem}, which states that quantum mechanics cannot be fully explained by hidden variable models that assign predetermined values for measurement outcomes in a non-contextual way. Unlike Bell's-based scenarios, these inequalities do not rely on entanglement and are applicable to single-party quantum systems
~\cite{lapkiewicz2011experimental,um2013experimental, meng2025contextuality, Fabiocontext}. 

\begin{figure*}[t!]
    \centering
    \includegraphics[width=1\textwidth]{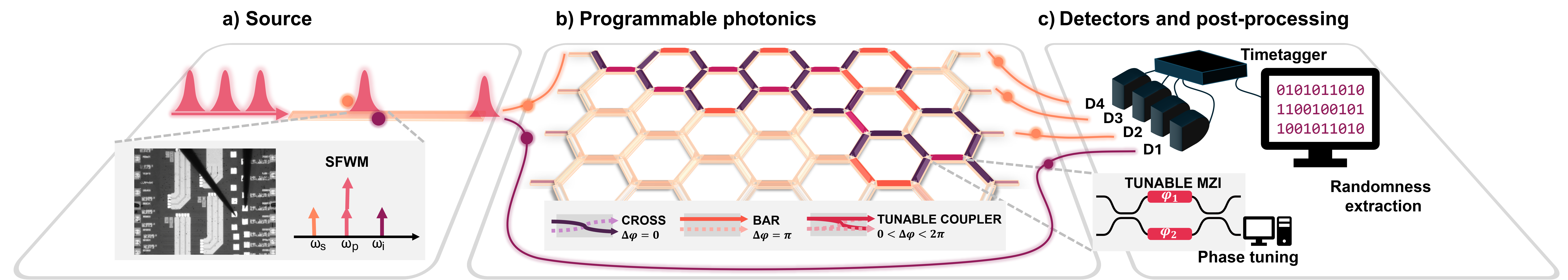}
    \caption{QRNG experimental scheme. \textbf{a) }Photon-pair source using spontaneous four-wave-mixing (SFWM). In the inset a picture of the source PIC. \textbf{b)} Programmable photonic circuit to manipulate single photons and perform qutrit rotations, consisting of 72 MZIs arranged in a hexagonal lattice. Each MZI can be reconfigured to perform different transformations of the two optical modes it operates on: the BAR operation ($\Delta\varphi=\pi$), the CROSS operation ($\Delta\varphi=0$), or any other tuneable two-mode coupler operation($0<\Delta \varphi<2\pi$), to realize a tuneable coupler $\Delta\varphi=\varphi_2-\varphi_1$. \textbf{c)} Detectors used to perform measurements: superconductive nanowire single-photon detectors (SNSPDs) and time taggers used for the counting logic. Randomness is extracted and certified in post-processing. More information about the experimental setup can be found in the Supplementary Section I. }
    \label{setup}
\end{figure*}
\noindent In this context, integrated silicon photonics has recently emerged as a promising platform for the realization of quantum technologies in a compact, stable and scalable way~\cite{Spring:13,silverstone2014chip,perez2017silicon,paesani2019generation, wang2018multidimensional}. Thanks to photonic integrated circuits (PICs), on-chip generation and manipulation of quantum states across multiple photonic degrees of freedom, including path, polarization, time-bin, and frequency encoding, enabling scalable architectures for quantum communication, computing, and sensing~\cite{borghi2023,congia25,kues2017chip} has been achieved.

Implementations of integrated QRNGs have been demonstrated, relying on diverse quantum processes such as homodyne detection~\cite{PRXQuantum.4.010330,raffaelli1612chip}, vacuum fluctuations~\cite{haylock2019multiplexed}, and photon arrival time~\cite{bisadi2018compact}. However, most of these schemes operate under fully trusted assumptions and do not provide certified randomness in the presence of tampered, imperfect or partially characterized devices. 

While contextuality-based randomness certification has been experimentally demonstrated in bulk optical systems, and contextuality witnesses have recently been certified on programmable integrated photonic processors \cite{giordani2023experimental}, composable randomness extraction from contextuality within integrated platforms had not previously been demonstrated.

\noindent By combining the advantages of integrated silicon photonics with contextuality-based certification, we demonstrate a proof-of-concept integrated, semi-DI QRNG. 
Our work establishes an end-to-end photonic architecture for contextuality-certified randomness generation, integrating photon generation, programmable state manipulation, contextuality verification, and randomness extraction within silicon photonic platforms.

Randomness is certified via the violation of the Klyachko-Can-Binicioğlu-Shumovsky (KCBS) inequality~\cite{PhysRevLett.101.020403, lapkiewicz2011experimental,um2013experimental}, demonstrating contextuality-based randomness generation in integrated photonic chips. Our scheme combines two different PICs: a heralded single-photon source and a reconfigurable interferometer mesh to prepare and measure quantum states. The KCBS scenario represents the simplest case in which non-contextual hidden variable (NCHV) models fail, namely a high-dimensional system with a Hilbert space dimension of three: a qutrit. By placing limits on the correlations allowed in a non-contextual theory, the KCBS inequality provides a minimal and experimentally accessible test of quantum contextuality, suitable to certify QRNGs. 

An observed violation of $-3.84 \pm 0.08$, against the classical bound of $-3$, confirms the presence of contextuality. We use this result to certify a conditional min-entropy per experimental round of $H_{min}=0.077\pm0.002$, obtained from a bespoke security analysis based on semidefinite programming. This guarantees that each measurement outcome contains at least $0.077\pm0.002$ bits of genuine randomness available for extraction, leading to an asymptotic rate of $21.7\pm0.5$ genuine random bits per second. 

We emphasize that the present implementation is not a randomness-expansion protocol. The measurement settings are selected using an independent trusted random seed, as is standard in certified randomness protocols, and the certified randomness is obtained from post-selected accepted measurement rounds. The goal of the present work is therefore to demonstrate randomness certification within a semi-device-independent framework, rather than positive net randomness expansion.
\noindent Although the present implementation is primarily intended as a proof-of-principle demonstration rather than an optimized high-throughput generator, it establishes the compatibility between contextuality-based certification protocols and programmable integrated photonic hardware.
Our integrated semi–DI QRNG leverages the compactness of photonic integrated circuits, significantly reducing the footprint of the overall quantum system.
At the same time, it opens a path towards implementing contextuality-based certification protocols within integrated quantum photonic technologies, such as photonic quantum key distribution (QKD) systems~\cite{euroqci2025,huttner2022long} or distributed quantum computing platforms, all based on single-photon transmission through fiber or free-space channels across distant nodes~\cite{Wehner:QuantumInternet}.

\section{Results}
\subsection{QRNG experimental scheme}

To realize a semi-DI QRNG in the KCBS scenario, five cyclic, pairwise compatible dichotomic measurements on a single qutrit system must be implemented, in order to verify a contextual behaviour in the system~\cite{PhysRevLett.101.020403}. The measurements correspond to specific projectors in a three-dimensional Hilbert space and must satisfy compatibility conditions that allow contextuality to emerge. Such verification imposes stringent requirements on the level of control over the quantum state preparation and the measurements, making integrated photonic chips with reconfigurable meshes ideal candidates for their implementation.

\noindent The proposed experimental setup (Fig.~\ref{setup}) consists of three components: an integrated heralded single-photon source (Fig.~\ref{setup}.a), an integrated programmable silicon photonic Mach-Zehnder-interferometer (MZI) mesh~\cite{perez2017silicon} to implement qutrit operations~\cite{universaloptics} (Fig.~\ref{setup}.b), and high-efficiency superconducting nanowire single-photon detectors (SNSPDs) connected to a time tagger for photon counting measurements, followed by randomness extraction in post processing (Fig.~\ref{setup}.c).

\begin{figure*}[ht!]
    \centering
    \includegraphics[width=1\textwidth]{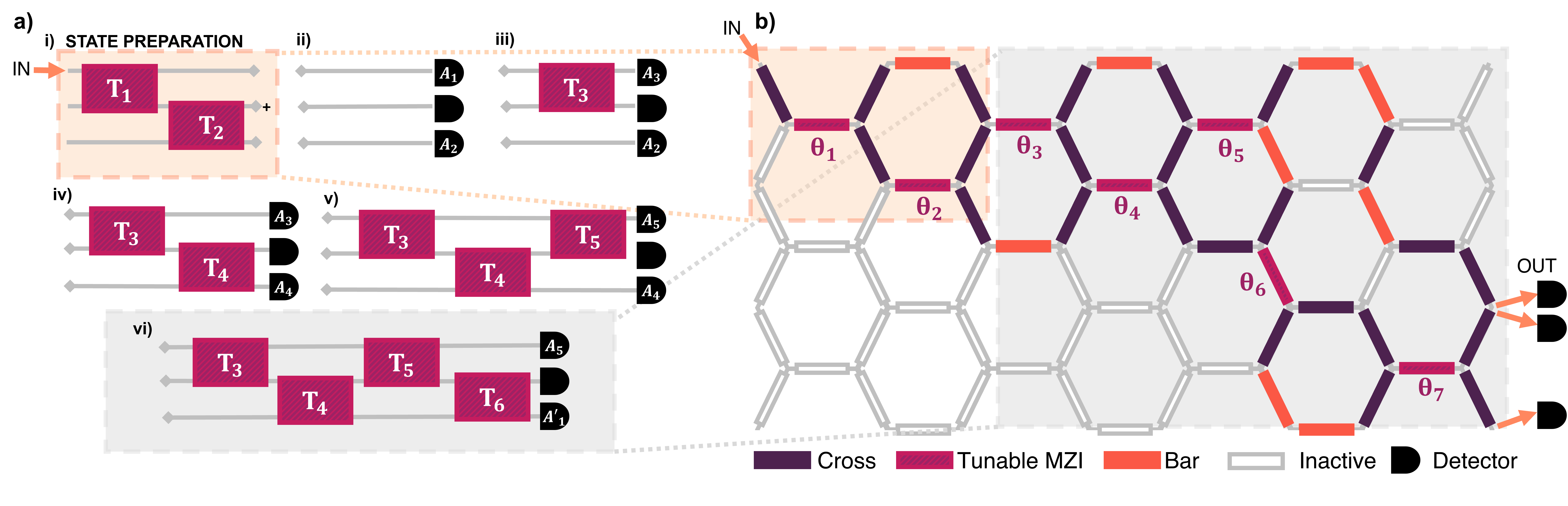}
    \caption{Experimental scheme implementation. \textbf{a)} Full scheme for the testing of the KCBS inequality. Grey lines represent optical modes; $T_i$ refer to the transformations performed on pairs of optical modes to implement the five measurement contexts. First, the desired input state is prepared in the i) step. Then, i) is followed by one of the contexts from ii)-vi). Photons are injected into the IN mode and collected at the output through single-photon detection. A coincidental detection between any of the three detectors placed on the three optical modes defining the qutrit space and the heralding photon, determines the measurement outcome of the two observables for each of the different contexts needed for building the inequality. \textbf{b)} Mapping of required transformations onto the hexagonal MZI mesh. $\theta$s can be directly mapped to the $T_i$ on the left. More information can be found in the Supplementary Section II. }
    \label{expdes}
\end{figure*}

The KCBS inequality considers five dichotomic observables $A_i$ ($i = 1, 2, 3, 4, 5$), each with possible outcomes $\pm1$. These observables are arranged in a cyclic compatibility structure, such that each $A_i$ is compatible (i.e., jointly measurable) with $A_{i+1}$. In any NCHV
theory, the expectation values $\langle A_i A_{i+1} \rangle$ must satisfy the inequality~\cite{PhysRevLett.101.020403}:
\begin{equation}
\begin{split}
\chi_{\text{KCBS}}= & \langle A_1 A_2 \rangle + \langle A_2 A_3 \rangle + \langle A_3 A_4 \rangle + \\ &  \langle A_4 A_5 \rangle + \langle A_5 A_1 \rangle \geq -3.
\end{split}
\label{kcbsin}
\end{equation}
These compatibility relations form the well-known KCBS pentagram graph, where each observable shares a measurement context only with its two neighbors.
This inequality provides a state-dependent test for contextuality by setting a limit on the correlations that can be observed in case the system were non-contextual.
For the canonical pentagram configuration of the observables \(A_i\), the inequality attains its maximal violation for a state \(|\psi_0\rangle\) that lies along the symmetry axis of the pentagram in the in Hilbert space. In our implementation, it can be expressed as
\begin{equation}
|\psi_0\rangle = \frac{1}{\sqrt[4]{5}} \, |0\rangle + \frac{1}{\sqrt[4]{5}} \, |1\rangle + \sqrt{1 - \frac{2}{\sqrt{5}}} \, |2\rangle ,
\label{eq:psi_0}
\end{equation}
leading to the minimal quantum expectation value of the KCBS expression,
\begin{equation}
\langle \mathrm{KCBS} \rangle_{\psi_0} = 5 - 4\sqrt{5} \approx -3.944,
\end{equation}
which surpasses the noncontextual bound~\cite{PhysRevLett.101.020403}. More information on the construction of the observables $A_i$ can be found in the Supplementary Information, Section II. 
A KCBS inequality violation indicates that there is no joint probability distribution that can account for the observed outcomes, thus, confirming the contextual nature of quantum mechanics. Therefore, measurement outcomes cannot be predetermined by any NCHV model. This certifies irreducible unpredictability that can be quantified as min-entropy of the system. To quantify the latter, we treat each experimental trial as a \textit{round}, defined as a single pump pulse irrespective of whether it produces a single-photon detection. In each round, one of the KCBS measurement contexts is selected, and the corresponding dichotomic observables \(A_iA_{i+1}\) is measured.
This intrinsic randomness can be quantified through the min-entropy \(H_{\min}\), which characterizes an adversary’s optimal guessing probability for the measurement outcome. For a given observed KCBS value \(\chi_{\mathrm{KCBS}}\), a bound on \(H_{\min}\) can be derived. More information on the min-entropy estimation can be found in the Supplementary Information, Section V and VII.
In our implementation, we extract a raw random bit in each round by assigning the two possible coincidence-detection outcomes of the measured observables \(A_iA_{i+1}\) to the bit values \(\{0,1\}\). After this mapping, a Toeplitz extractor~\cite{hash} is applied to the data using the experimentally estimated min-entropy bound, yielding a final secure random bit string certified by the observed KCBS violation.

\subsection{On-chip implementation and violation of the KCBS inequality}

In order to initialize the qutrit \(| \psi_0\rangle \), we generate photon pairs through spontaneous four-wave mixing (SFWM) in a silicon waveguide. Here, the interaction between two pump photons and the silicon waveguide produces a pair of non-degenerate signal and idler photons~\cite{Spring:13,silverstone2014chip}. The chip used for the photon-pair generation was fabricated in a SiGe-BiCMOS line at IHP - Leibniz Institute for High Performance Microelectronics~\cite{Gajda:12, doi:10.1126/sciadv.adk6890} and included a p-i-n junction along the waveguide to decrease free-carrier absorption.
Photon pairs were generated by pumping a $2$-cm long waveguide with 5-ps laser pulses at a $1.25$~GHz repetition rate, and $\approx6$~dBm of average power, providing a pair generation efficiency around $3\%$. Detection of the idler photon by a SNSPD heralds the presence of a signal photon, which in turn can be processed by linear optical elements to encode a qutrit in its path degree of freedom. The heralded signal photon is then injected into a second PIC containing a reconfigurable hexagonal interferometer mesh~\cite{perez2017silicon}, which can implement arbitrary unitary transformations up to six waveguide modes. This architecture allows us to prepare the desired qutrit \(| \psi_0\rangle \) and realize the sequence of five dichotomic projective measurements required to test the KCBS inequality. At the output of the programmable mesh, the transformed single photon is directed to three additional SNSPDs. Selected coincidental detection events are used to reconstruct the measurement statistics, both to evaluate the degree of contextuality and to build the random sequence. The observed violation is then used to certify the generation of random bits in a semi-DI framework. In this model, the source is treated as untrusted, while the measurement device and acceptance rule are assumed to be trusted, and the measurement settings are selected independently. In addition, the certification incorporates experimentally calibrated parameters, including measurement compatibility, overlap between observables, detected-round statistics, and multiphoton contributions.

\noindent The measurement scheme, designed to test the inequality, is shown in Fig.~\ref{expdes}.a and consists of six main steps. First, the state is prepared by creating a superposition of the input photon among three waveguide modes (Fig.~\ref{expdes}.a.i). After the state preparation, step a.i is then followed by one of the five measurement contexts ii-vi. At each context, the modes are being mixed by the transformation $T_i$. A key aspect is that each transformation acts only on two modes, leaving the remaining mode unaffected (e.g., the part of the physical setup corresponding to the measurement of $A_2$ is the same in Fig.~\ref{expdes}.a.ii and \ref{expdes}.a.iii). The response of two detectors monitoring the optical modes defines a pair of measurement outcomes as appearing in the KCBS inequality (Eq.~\ref{kcbsin}). The outcomes of the measurements ($A_i$) are determined by asking whether the detector has registered a detection or not. A detection (no-detection) event results in a value of -1 (+1) assigned to the corresponding measurement.

\noindent In our implementation, the three-level qutrit system is encoded in three distinct optical paths, defined by configuring the interferometer mesh accordingly (Fig.~\ref{expdes}.b). The transformations $T_i$ between the modes are implemented by tunable MZIs. The heralded single photons are coupled into the first mode (IN in Fig.~\ref{expdes}.b), with the initial qutrit state prepared via transformations $T_1$ and $T_2$, implemented via MZI($\theta_1$) and MZI($\theta_2$) respectively. The five measurement contexts required for the KCBS inequality were implemented by reprogramming the photonic circuit into five experimental configurations, corresponding to the five measurement contexts (the final configuration is shown in Fig.~\ref{expdes}.b). Each MZI-trasformation mixes only two modes at a time in the active regions (highlighted in Fig.~\ref{expdes}.b with \( \theta_i \)), while the unused sections of the mesh are set to implement an overall identity operation on each mode. This approach enables the realization of the required measurement contexts for the KCBS scenario. 
\begin{figure}
    \centering
    \includegraphics[width=1\linewidth]{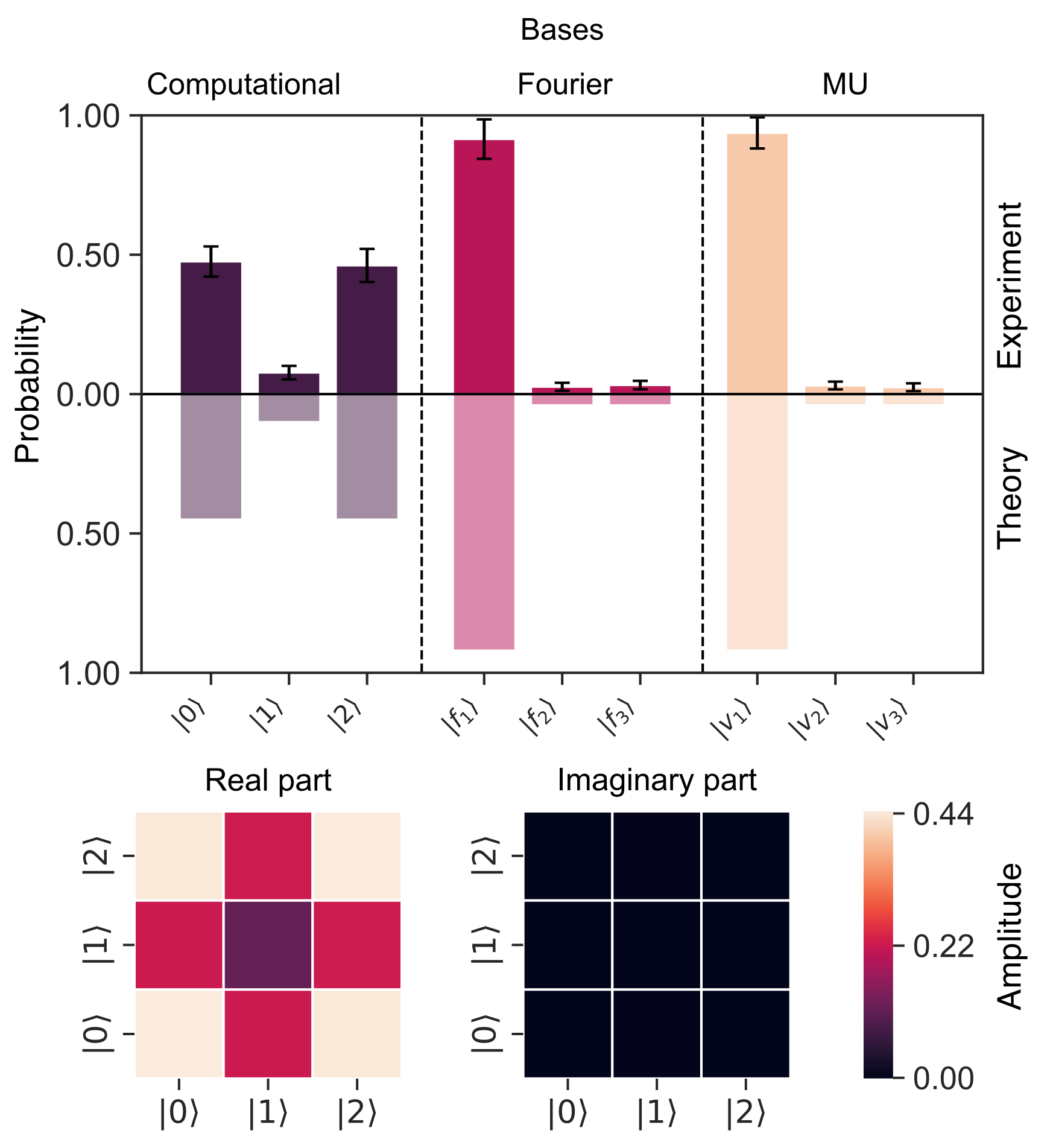}
    \caption{\textbf{Top.} Measured outcome probabilities for quantum state tomography performed in three mutually unbiased bases: the computational basis $\{|0\rangle,|1\rangle,|2\rangle\}$, the Fourier basis $\{|f_0\rangle,|f_1\rangle,|f_2\rangle\}$, and a third mutually unbiased (MU) basis  $\{|v_0\rangle,|v_1\rangle,|v_2\rangle\}$. Bars show the experimental results with statistical uncertainties, and shaded regions indicate the ideal theoretical expectations. 
    \textbf{Bottom.} Reconstructed density matrix of the prepared qutrit state, showing the real and imaginary components. Uncertainties are computed assuming Poissonian statistics for the raw counts and then propagated.  More information regarding quantum state tomography can be found in the Supplementary Section III. }
    \label{denmat}
\end{figure}

\noindent To verify the correct initialization of the qutrit state $|\psi_0\rangle$ and the accurate implementation of the unitaries defining the measurement contexts, we performed quantum state tomography (QST) using a maximum-likelihood estimation (MLE) algorithm. The measured outcome probabilities in the three mutually unbiased bases are shown in Fig.~\ref{denmat}, and agree with the theoretical predictions within statistical uncertainty. Further details on the MLE procedure are provided in the Supplementary Section III. The reconstructed density matrix, also shown in Fig.~\ref{denmat}, exhibits a fidelity of $F = 0.99 \pm 0.01$ with the target state, confirming precise control of the operations implemented on the integrated photonic MZI mesh.
\begin{table*}[ht!]
\begin{tabular*}{\textwidth}{@{\extracolsep\fill}llclclclc}
\toprule
\hline
\multicolumn{2}{c}{D2} & \multicolumn{2}{c}{D3} & \multicolumn{2}{c}{D4} & \multicolumn{2}{c}{Expectation Value} \\
\hline
\cmidrule{1-2} \cmidrule{3-4} \cmidrule{5-6} \cmidrule{7-8}
Condition & Value & Condition & Value & Condition & Value & Term & Value \\
\midrule
$P(A_1 = -1, A_2 = +1)$ & $0.42\pm0.03$ & $P(A_1 = +1, A_2 = +1)$ & $0.11\pm0.02$ & $P(A_1 = +1, A_2 = -1)$ & $0.47\pm0.03$ & $\langle A_1 A_2 \rangle$ & $-0.78\pm0.04$ \\

$P(A_2 = -1, A_3 = -1)$ & $0.10\pm0.03$ & $P(A_2 = +1, A_3 = +1)$ & $0.45\pm0.02$ & $P(A_2 = +1, A_3 = -1)$ & $0.45\pm0.03$ & $\langle A_2 A_3 \rangle$ & $-0.80\pm0.04$ \\

$P(A_3 = +1, A_4 = -1)$ & $0.42\pm0.03$ & $P(A_3 = -1, A_4 = +1)$ & $0.47\pm0.02$ & $P(A_3 = +1, A_4 = +1)$ & $0.11\pm0.03$ & $\langle A_3 A_4 \rangle$ & $-0.77\pm0.03$ \\

$P(A_4 = -1, A_5 = +1)$ & $0.42\pm0.03$ & $P(A_4 = +1, A_5 = +1)$ & $0.11\pm0.02$ & $P(A_4 = +1, A_5 = -1)$ & $0.47\pm0.03$ & $\langle A_4 A_5 \rangle$ & $-0.78\pm0.04$ \\

$P(A_5 = +1, A_1 = -1)$ & $0.44\pm0.03$ & $P(A_5 = +1, A_1 = +1)$ & $0.11\pm0.02$ & $P(A_5 = -1, A_1 = +1)$ & $0.46\pm0.03$ & $\langle A_5 A_1 \rangle$ & $-0.79\pm0.04$ \\
\botrule
\end{tabular*}
\caption{Measured probabilities used to compute the KCBS expectation values. Each row corresponds to one term $\langle A_i A_j \rangle$ in the inequality, derived from joint probability measurements. Detector $D_1$ is used for heralding. Uncertainties are computed using standard error propagation, assuming a Gaussian distribution on the coincidence counts.}
\label{table:probabilities_contributions}
\end{table*}

\noindent A measurement period of $15$ minutes is chosen to gather the statistics of the heralded outcomes of all the 5 contexts shown in Fig. \ref{expdes}.a. The measured joint probabilities used to compute the KCBS expectation values are summarized in Table~\ref{table:probabilities_contributions}. The probabilities have been corrected for differences in detector efficiencies. The expectation values $\langle A_i A_j \rangle$ are computed as:
\begin{equation}
\langle A_iA_j\rangle=\mathbb{P}(+,+)+\mathbb{P}(-,-)-\mathbb{P}(+,-)-\mathbb{P}(-,+).
\end{equation}
Conditioned on accepted single-click rounds, the auxiliary detector implements the $(-,-)$ outcome in the corresponding KCBS context, while no-click and multi-click events are rejected by the acceptance rule. The dichotomic observables can be implemeneted as:
\begin{equation}
    A_i=I-2\,\Pi_i,\;\text{where} \;\Pi_i=|\psi_i\rangle\langle\psi_i|, \; \text{and} \;\Pi_i\Pi_{i+1}=0,
\end{equation}
with ideal case $[A_i,A_{i+1}]=0$. A detection event in each projector $\Pi_i$ yields $A_i=-1$; otherwise $A_i=+1$.
It is important to note that $A_1$ and $A'_1$, see Fig. \ref{expdes}.ii and Fig. \ref{expdes}.vi, are not identical, and hence strictly violate the cyclic compatibility condition on which the KCBS inequality is defined. 
To solve this problem, similarly to Ref.~\cite{lapkiewicz2011experimental,um2013experimental}, we use a modified inequality that includes the $A'_1$ observable with the form:
\begin{equation}
\begin{split}
\chi'_{\text{KCBS}}= & \langle A_1 A_2 \rangle + \langle A_2 A_3 \rangle + \langle A_3 A_4 \rangle + \langle A_4 A_5 \rangle\\ + &  \langle A_5 A'_1 \rangle + [1-\langle A_1 A'_1 \rangle]\geq -3.
\end{split}
\label{kcbsinmod}
\end{equation}

\noindent Note that inequality (\ref{kcbsinmod}) becomes the original KCBS inequality in eq. (\ref{kcbsin}), when $A_1=A'_1$. Therefore, the difference between the two measurements, quantifiable by the overlap parameter $R = \langle A_1 A'_1 \rangle$, i.e. the overlap between the two measurements $A_1, A'_1$,  decreases the value of the maximum violation that can be experimentally obtained.  More details on the compatibility loophole can be found in the Supplementary Section IV.
We report a measured value of $\chi'_{\mathrm{KCBS}}=-3.84 \pm 0.08$, having estimated the overlap parameter $R = 0.93 \pm 0.01$. Uncertainties are computed assuming Poissonian statistics for the raw counts and then propagated.  This value violates the classical bound of $-3$ by over $10$ standard deviations and confirms the presence of contextual behaviour in our single-qutrit implemented on-chip, ruling out NCHV models. 
This violation provides the core resource for certifying quantum randomness in our semi-DI QRNG: the measured photon detection events both certify randomness through the violation of the KCBS inequality above, and also provide the raw sequence to extract the final certified random string of bits. More specifically, we link each signal-idler coincidence event on the modes that define the qutrit to the individual dichotomic observables outcomes of the KCBS inequality $\langle A_i \rangle \in \{+1, -1\}$. For every context measured, the three qutrit spatial modes are mapped to two QRNG modes and one auxiliary mode, denoted by $m_i$, $m_{i+1}$ and $m_{\mathrm{aux}}$. Then we assign bit values \(\{0,1\}\) to the two possible coincidence-detection outcomes on modes $m_i$, $m_{i+1}$, forming the raw binary random sequence.

\subsection{Extractable randomness}

\begin{figure}[t!]
    \centering
    \includegraphics[width=0.8\linewidth]{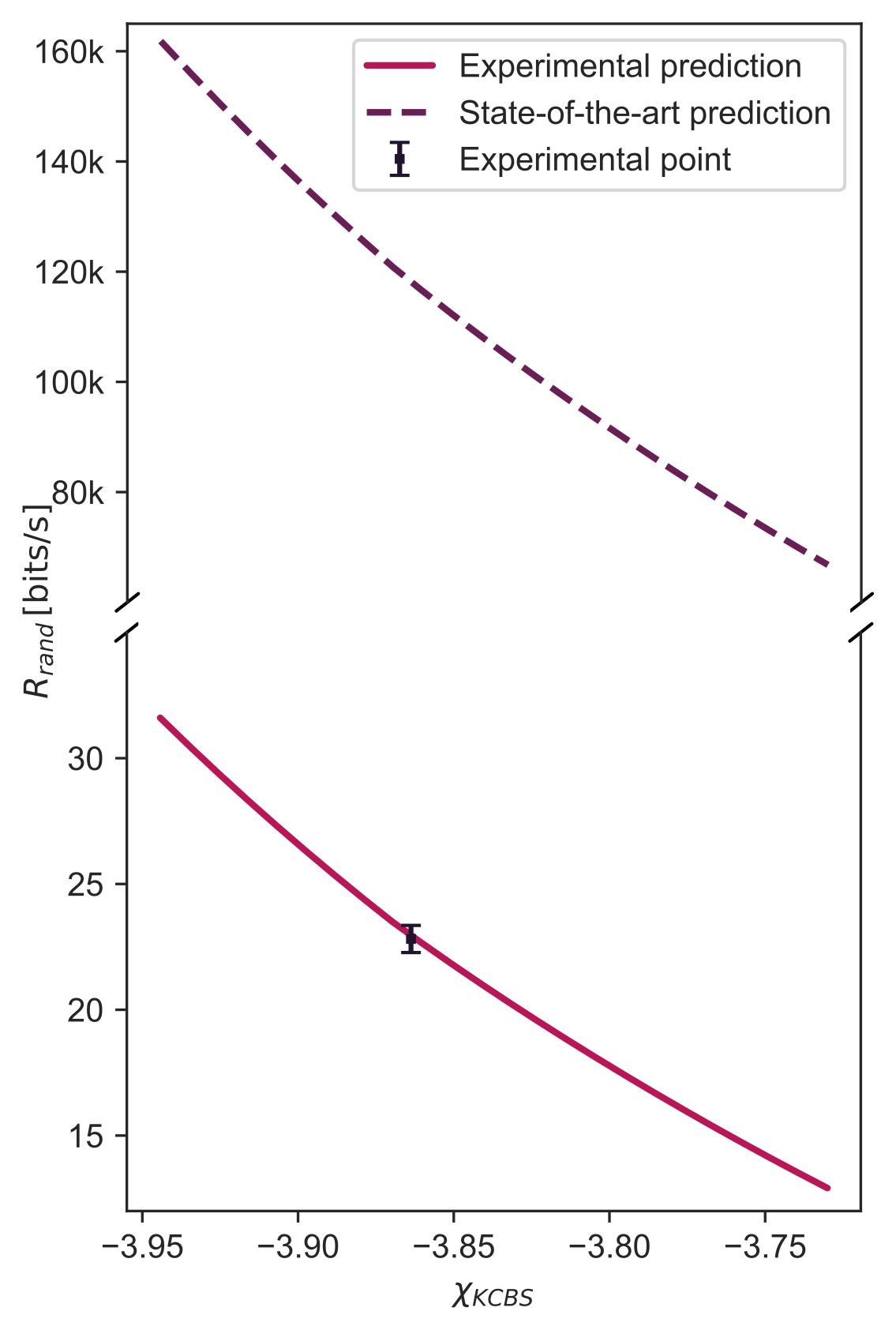}
    \caption{Certified quantum random number generation rate as a function of the value of the KCBS witness $\chi_{\mathrm{KCBS}}$. The solid line represents the theoretical prediction including the experimentally calibrated device imperfections, detector inefficiencies, and chip losses. The experimental point with vertical error bar certifies $0.077\pm0.002~\mathrm{bits/round}$, which leads to $R_{\text{rand}} = 21.7\pm0.5~\text{bit/s}$, in good agreement with the theoretical prediction. The dashed line shows the theoretical prediction in the ideal experimental case, i.e. with state-of-the-art SFWM sources, chip losses, and detectors.}
    \label{Fig::H_min_vs_KCBS}
\end{figure}

The randomness generated by our semi-DI QRNG is certified through a security analysis based on semidefinite programming (SDP). In this framework, the amount of extractable randomness is directly related to the observed violation of the KCBS inequality, without requiring full device characterization or assumptions on the internal workings of the measurement apparatus, except for the dimension of the Hilbert space. 
We model the adversary’s knowledge as the quantum side information Eve ($E$) might have access to, when entangled with the system. The relevant figure of merit is the conditional min-entropy $H_{\text{min}}(A|E)$, which quantifies the unpredictability of the measurement outcomes $A=\{A_i\}_{i=1,..,5}$ given access to $E$. To bound this quantity, we construct an SDP model that maximizes the guessing probability $P_{\text{guess}}(A|E)$ over all quantum states and measurements compatible with the observed KCBS violation:
\[
H_{\text{min}}(A|E) = -\log_2 P_{\text{guess}}(A|E).
\]
The overall certification follows a standard proof chain, in which the SDP provides an upper bound on the adversarial guessing probability, which is then converted into a single-round min-entropy, extended to finite size via entropy accumulation, and finally converted into uniform random bits through quantum-proof extraction.
The SDP constraints are derived from the contextuality correlations and the quantum dimension assumption (qutrit), ensuring that the optimization remains within physically valid states and measurements.
Figure~\ref{Fig::H_min_vs_KCBS} shows the variation of the QRNG rates, and therefore the certified $H_{\min}(A|E)$ linked to it, as a function of the violation of the KCBS inequality.
Each point on the curve corresponds to a solution of the SDP with consideration on Eve's measurement and operator confinements, calibrated near-commutation $\|[A_i, A_{i+1}]\|_\Gamma \le \epsilon_{\mathrm{com}}$ with $\epsilon_{\mathrm{com}} = 0.047$ defined in the Supplementary Information, Section VII, and the overlap constraint for the modified KCBS inequality, $R = \langle A_1 A'_1 \rangle \in [0.92, 0.94]$, given the experimentally obtained value of $R_{comp} = 0.93 \pm 0.01$ (Supplementary Information, Section IV). The experimental worst-case point corresponds to $\hat{\chi}_{\text{KCBS}} = -3.92$ with finite-size effect due to $n_i = 10^5$ detected rounds, which leads to a Hoeffding deviation of $\Delta = 0.056$. The SDP constraints also include the measured detection efficiencies of the three detectors situated on each spatial mode: $\boldsymbol{\eta} = (\eta_1, \eta_2, \eta_3) = (0.85, 0.83, 0.86)$. Under these constraints, and given our KCBS experimental results, we find a conditional min-entropy value of $H_{\min}(A|E)=0.077\pm0.002$, which corresponds to the number of extractable certified bits for each round of measurements in the QRNG routine, where we assume that Alice selects a stage at random in each round and performs single-shot photon readout. More information about the estimation of $H_{\min}(A|E)$ can be found in the Supplementary Information Section V and VII.
To extract certified uniform random bits from the raw measurement outcomes, we apply a quantum-proof randomness extractor based on Toeplitz matrix hashing~\cite{hash}. The extractor compresses blocks of raw data into a smaller set of nearly uniform bits, secure against quantum adversaries, using the above $H_{\min} (A|E)$. The final randomness generation rate, after correcting for finite-size effects, detector efficiencies, and security parameters, is therefore $R_{\text{rand}} = 21.7\pm0.5~\text{bit/s}.$ Randomness is further validated through the NIST SP 800-22 suite (Supplementary Information, Section VI).

\section{Discussion}

Semi-DI QRNGs represent a pragmatic compromise, providing meaningful security rooted in quantum principles while remaining compatible with scalable, integrated chips and avoiding the demanding entanglement resources and large Hilbert spaces required for fully DI, Bell-based implementations.

In particular, contextuality-based protocols are well suited to photonic implementations, as they operate on single quantum systems and therefore avoid the need for distributed entanglement while still enabling certified randomness under relaxed trust assumptions.

\noindent The assumption behind the KCBS inequality is that the observables $A_i$ and $A_{(i+1)}$ are compatible. However, verifying compatibility of observables is not straightforward in an actual experiment, which leads to the opening of the compatibility loophole. This issue has been addressed by modifying the KCBS inequality, according to previous works~\cite{PhysRevA.81.022121,PhysRevLett.130.080802}. Further details are provided in the Supplementary Information, Section IV. Alternative measurement schemes can be used to perform cleaner contextuality tests which naturally avoid such compatibility loophole~\cite{TerraCuna_Bell_Contex}.

\noindent The observed violation certifies a nonzero min-entropy per round, bounded directly by the degree of KCBS violation. 
In our current implementation, the interferometric mesh exhibits an overall optical loss of $\sim 27$~dB, leading to a modest random-bit generation rate of $\sim 22$~bits/s. 

\noindent This rate should be interpreted as a proof-of-principle demonstration rather than a technological limit of contextuality-based QRNGs.

\noindent These losses are dominated by in- and out-coupling of the PIC. Reducing the total loss to 14~dB—achievable with state-of-the-art couplers~\cite{Ding:14}-would increase the transmission by nearly a factor of 20. Such improvements are well within reach using existing photonic foundry processes, when embedded in photonic devices for more advanced computing and communication systems.

\noindent Looking ahead, our results establish a new path toward chip-scale semi-DI QRNGs with enhanced performance. Higher rates can be pursued with the monolithic integration of photon sources and interferometric circuits, and by moving from a general-purpose programmable architecture to dedicated mesh designs for contextuality-based protocols. 

\noindent In our current setup, the dominant loss arises from in- and out-coupling between separate chips, including $\sim 7$~dB of source-to-mesh coupling. Removing these interfaces avoids additional input coupling loss. For a dedicated rectangular $7\times 7$ mesh with propagation losses of $0.5$~dB per MZI, low-loss output couplers ($\sim 0.6$~dB)~\cite{Ding:14}, and SNSPD operating at $\sim95\%$ efficiency, the overall transmission can exceed $30\%$. Combined with heralded photon-pair sources on silicon with ideal purity reaching $1.5$~MHz~\cite{bright}, this performance would enable random-bit generation rates on the order of $120$~kbits/s (see dashed line in Fig.~\ref{Fig::H_min_vs_KCBS}).

\noindent This projection is obtained within the same semi-device-independent certification framework and security model used in the present experiment, and should be understood as a model-based estimate rather than a demonstrated operating regime.

\noindent Additional improvements could arise from multiplexed photon sources, reduced propagation losses in optimized waveguide platforms, and faster detection electronics, all of which are active areas of development in integrated quantum photonics.

\noindent Furthermore, the ability to verify contextuality on-demand within a reprogrammable photonic hardware establishes a foundation for future quantum information tasks based on state-dependent contextuality, including semi–DI QKD \cite{Review_DI_QKD} and randomness amplification \cite{kessler2020device, Colbeck_2011_RandomnessExpansion}. 

\noindent More broadly, programmable photonic circuits capable of implementing contextuality tests may provide a versatile platform for exploring foundational quantum correlations alongside practical certification protocols within integrated hardware.

\noindent Finally, the inherently distributed nature of our two-chip system enables QRNG implementations that can be seamlessly integrated into more complex networked quantum platforms, such as photonic quantum communication networks and distributed photonic quantum computing systems, where the processing nodes manipulating distributed quantum resources may be untrusted. 

\noindent Crucially, this functionality can be achieved without requiring additional physical resources or dedicated photonic components for the QRNG, thereby avoiding any hardware overhead.

\section{Acknowledgments}
This project has received funding from the European Union’s HORIZON-CL4-2021-DIGITAL-EMERGING-01 programme under the PROMETHEUS project (Grant Agreement No. 101070195), and from the European Union’s Horizon Europe Research and Innovation Programme under the QPIC 1550 project (Grant Agreement No. 101135785). This work was also supported in part by the German Research Foundation (DFG) through the project ULTRA2 (Grant Agreement No. ZI~1283/5-2) within the Priority Programme SPP~2111. F.D.R. acknowledges financial support from the Villum Foundation ( project OPTIC-AI grant n. VIL29344). D.B. acknowledges support from the European Union ERC StG, QOMUNE, 101077917.

The authors also thank Battulga Munkhbat for providing access to the experimental resources, Emanuele Polino and Yu Meng for insightful discussions on quantum contextuality. Daniel Perez and the iPronics team are acknowledged for technical support with Smartlight. Claudio Pereti and Giacomo Ferranti from QTI are acknowledged for insightful discussion on randomness extraction.

\section{Methods}
\textsmaller{

\subsection*{Photonic interferometer mesh}
The inequality testing was implemented on a silicon-based programmable photonic processor (iPronics Smartlight)~\cite{rausell2024universal, perez2017silicon}, composed of 72 programmable unit cells (PUCs) arranged in a hexagonal mesh. Each PUC comprises a tunable Mach-Zehnder interferometer (MZI) with internal thermo-optic phase shifters $\varphi_1, \varphi_2$ on each arm, allowing arbitrary SU(2) transformations between pairs of modes. This allows full control of both an induced relative phase between the two modes $\Delta^- \varphi= (\varphi_2- \varphi_1)/2$, as well as a global phase for both modes $\Delta^+ \varphi= (\varphi_2+\varphi_1)/2$ .

\subsection{Loss}
The total system loss includes contributions from both the source and interferometer mesh chips. The heralded single-photon source is integrated on a silicon photonic chip with an insertion loss of approximately 11~dB, which includes -in- and out-coupling and propagation losses within the chip. Photons are then edge-coupled into a second chip containing the programmable interferometer mesh. This stage constitutes the largest loss—about 27~dB—dominated by in- and out-coupling inefficiencies at the chip interfaces, with additional contributions from propagation loss of $0.5$~dB/PUC. These cumulative losses primarily limit the overall count rate.

\subsection{Quantum state tomography via MLE}
Quantum state tomography was performed to verify the preparation of the qutrit state used in the contextuality experiment. We used a maximum likelihood estimation (MLE) procedure \cite{PhysRevA.64.052312}, to reconstruct the density matrix $\rho$ from a set of projective measurements in mutually unbiased bases. For qutrits, we measured the state in three bases: the computational basis, the Fourier basis, and a third mutually unbiased (MU) basis derived from the Fourier basis \cite{brierley2010mutuallyunbiasedbasesdimensions}. 
The experimental data consisted of nine averaged coincidental photon count measurements, summarized in Fig.~\ref{denmat}. These were used to define a likelihood function, which was numerically optimized over the parameters needed to define the density matrix in the Gell-Mann representation. All measurements were averaged over 10-minute acquisition intervals. The statistical uncertainty in the density matrix elements was estimated assuming Poissonian count statistics. Details of the basis definitions, MLE derivation, optimization procedure and the reconstructed density matrix are presented in the Supplementary Information, section III.

The fidelity between the experimentally reconstructed quantum state \(\rho_{\mathrm{exp}}\) and the ideal target state \(\rho_{\mathrm{th}}\), used to quantify the accuracy of state preparation, is defined as \(F(\rho_{\mathrm{exp}}, \rho_{\mathrm{ideal}}) = \left(\mathrm{Tr} \left[ \sqrt{\sqrt{\rho_{\mathrm{ideal}}} \rho_{\mathrm{exp}} \sqrt{\rho_{\mathrm{ideal}}}} \right] \right)^2\). All fidelities reported in this work are calculated using this definition, based on density matrices reconstructed via maximum likelihood estimation. 

\subsection{Observables $A_i$}

The five observables entering the KCBS inequality correspond to projective measurements onto five qutrit states $\{|\psi_i\rangle\}_{i=1}^5$, embedded in a three-dimensional Hilbert space spanned by the optical path basis $\{|0\rangle,|1\rangle,|2\rangle\}$. 
Each observable is defined as
\begin{equation}
    A_i = \mathbb{I} - 2\,|\psi_i\rangle\langle\psi_i|,
\end{equation}
so that the outcome $-1$ corresponds to projection onto $|\psi_i\rangle$, while $+1$ corresponds to projection onto its orthogonal subspace. 
The KCBS compatibility condition,
\begin{equation}
    \langle \psi_i | \psi_{i+1} \rangle = 0 \qquad (i = 1, \dots, 5 \ \mathrm{mod}\,5),
\end{equation}
ensures that neighboring observables can be jointly measured.

In our implementation, the input qutrit state is a coherent superposition of the three optical paths,
\begin{equation}
    |\psi_{\mathrm{in}}\rangle =
    \frac{1}{\sqrt[4]{5}}\,(|0\rangle + |1\rangle)
    + \sqrt{1 - \frac{2}{\sqrt{5}}}\,|2\rangle,
\end{equation}
which can be viewed as a rotated version of the ideal KCBS state $|2\rangle$ in the canonical theoretical frame. 
Accordingly, the measurement states $|\psi_i\rangle$ are obtained by applying the same unitary rotation $U$ to the canonical KCBS vectors $|v_i\rangle$:
\begin{equation}
    |\psi_i\rangle = U\,|v_i\rangle, \qquad
    A_i = \mathbb{I} - 2\,|\psi_i\rangle\langle\psi_i|.
\end{equation}

\subsection{SDP certification}

The adversarial guessing probability underlying the randomness certification is computed via a semidefinite program formulated in a moment-matrix representation. The decision variable is a positive semidefinite matrix $\Gamma$, whose entries correspond to expectation values of operator products evaluated on the post-selected state $\rho^{(1)}_{AE}$. The operator set includes the observables $\{A_i\}$, $A'_1$, Eve’s operators $\{E_i\}$, and their relevant products, capturing both contextuality correlations and adversarial side information.

The feasible set of the SDP is defined by affine constraints derived from the experimental data and the semi-device-independent trust model. These include normalization, bounded single-body expectation values, interval constraints for KCBS-compatible correlators, the experimentally calibrated overlap parameter $R = \langle A_1 A'_1 \rangle$, and near-commutation conditions enforcing approximate compatibility between adjacent observables. Additional constraints incorporate detected-round statistics and bounds on multiphoton contributions. 

The objective function maximizes Eve’s average guessing probability over measurement contexts, taking into account the experimentally determined detection weights. The resulting optimal value provides an upper bound on the adversarial success probability, which is converted into a single-round conditional min-entropy via $H_{\min} = -\log_2 P^{\max}_{\mathrm{guess}}$. This quantity serves as the input to the finite-size entropy accumulation analysis and subsequent quantum-proof randomness extraction. The full SDP formulation and its numerical implementation are detailed in the Appendix Section VII.

\subsection{Experimental imperfections}

The randomness certification explicitly incorporates relevant experimental imperfections within the semi-device-independent framework. In particular, we account for imperfect compatibility between sequential measurements through experimentally calibrated near-commutation constraints, and for non-ideal overlaps between observables via the measured parameter $R = \langle A_1 A'_1 \rangle$. The analysis further includes bounds on multiphoton contributions and uses the observed detected-round statistics to weight the measurement outcomes. These quantities enter as affine constraints in the moment-matrix SDP, ensuring that the certified entropy remains valid under realistic deviations from the ideal KCBS scenario.

\subsection{Entropy accumulation and randomness extraction}

The single-round conditional min-entropy obtained from the SDP is extended to a finite-length setting using entropy accumulation techniques, which provide a composable lower bound on the total smooth min-entropy of the output string. The bound depends on the observed KCBS violation, the detected-round statistics, and the chosen security parameters. The resulting entropy is then converted into a uniformly random bit string using a quantum-proof randomness extractor, ensuring security against adversaries holding arbitrary quantum side information. This yields a final output string with a composable security guarantee quantified by the overall failure probability.

}

\bibliography{bib.bib}
\bibliographystyle{ieeetr}

\setcounter{equation}{0}
\setcounter{figure}{0}
\setcounter{page}{0}

\end{document}